# Effect of connectivity on the carrier transport and recombination dynamics of perovskite quantum dot networks

David O. Tiede[1], Carlos Romero-Pérez[1], Katherine A. Koch[2], K. Burak Ucer[2], Mauricio E. Calvo[1], Ajay Ram Srimath Kandada[2], Juan F. Galisteo-López[1]*, Hernán Míguez[1]*.

[1]Instituto de Ciencias de Materiales de Sevilla (Consejo Superior de Investigaciones Científicas-Universidad de Sevilla), C/Américo Vespucio, 49, Sevilla, 41092, Spain.

[2] Department of Physics and Center for Functional Materials, Wake Forest University, 1834 Wake Forest Road, Winston-Salem, NC 27109, USA.

*E-mail: juan.galisteo@csic.es, h.miguez@csic.es

**Abstract**: Quantum dot (QD) solids are being widely exploited as a solution-processable technology to develop photovoltaic, light-emission, and photo-detection devices. Charge transport in these materials is the result of a compromise between confinement at the individual QD level and electronic coupling among the different nanocrystals in the ensemble. While this is commonly achieved by ligand engineering in colloidal-based systems, ligand-free QD assemblies have recently emerged as an exciting alternative where nanostructures can be directly grown into porous matrices with optical quality as well as control over their connectivity and hence charge transport properties. In this context, we present a complete photophysical study comprising fluence and temperature-dependent time-resolved spectroscopy to study carrier dynamics in ligand-free QD networks with gradually varying degrees of interconnectivity, which we achieve by changing the average distance between the QDs. Analysis of the photoluminescence and absorption properties of the QD assemblies, involving both static and time-resolved measurements, allows us to identify the weight of the different recombination mechanisms, both radiative and non-radiative, as a function of QD connectivity. We propose a picture where carrier diffusion, which is needed for any optoelectronic application and implies inter-particle transport, gives rise to the exposure of carriers to a larger defect landscape than in the case of isolated QDs. The use of a broad range of fluences permits extracting valuable information for applications demanding either low or high carrier injection levels and highlighting the relevance of a judicious design to balance recombination and diffusion.

The potential of semiconductor QDs for optoelectronic applications is dictated by the electronic coupling between them when assembled into device compatible QD solids. Here, beyond the possibility of tuning the optical and electronic properties of individual QDs by acting on their size, composition, and ligand structure, one can also exploit the properties arising from collective interactions.[1] Controlling the QD separation, effective electronic and energy transport in devices can be attained while maintaining the characteristic quantum size confinement features. This has in fact already resulted in impressive advances in photovoltaic [2] as well as electrically pumped solution-processed lasers.[3] Within this context, over the past decade, lead halide perovskite QDs have emerged as an exciting alternative in the field of QD optoelectronics. [4] Together with the possibility of tuning their electronic properties by controlling their size, halide perovskite QDs have an additional degree of freedom in their composition where employing a combination of two halides (Cl/Br or Br/I) allows for tuning their electronic bandgap across the visible and near infrared spectral regions. [5,6] In this direction perovskite optoelectronic devices containing QD films as active layers, such as solar cells [7] LEDs [8] or photodetectors, [9] have recently demonstrated outstanding performances in line with or surpassing other semiconductor QD-based devices. As with the well-established II-VI and III-V inorganic semiconductor QDs, colloidal synthesis in its various forms [4] has become the main synthetic route for perovskite QDs. Within this approach, interparticle spacing could be tuned by varying the ligand employed,[10, 11] which adds complexity to the analysis of collective effects, since each ligand may give rise to different transport properties.[12]-[14] Alternatively, confinement can be achieved by synthesizing QDs within porous matrices, where the interparticle distance is controlled by the concentration of nanocrystals, which is in turn determined by the porosity of the host matrix and the precursor concentration [15-20]. This approach permits the synthesis of highly emitting and stable perovskite QDs directly into thin films with optical quality amenable to being introduced as an active layer within an optoelectronic device. [17,21] Central to the analysis presented herein, these nanostructures allow studying the photophysical properties of nanostructures isolated from any ligand or solvent-induced effect. [22]

In this work, we employ ligand-free perovskite QDs embedded within the pores of metal-oxide nanoporous films as a test bench to understand the different collective phenomena leading to charge carrier recombination in a QD solid. By tuning the average inter-QD separation we are able to controllably modulate the interaction between them and move from a scenario of isolated QDs to one of interconnected QDs where collective phenomena dictate the optoelectronic properties of the network. Time-resolved spectroscopic techniques are employed to study charge carrier recombination and unveil the gradually increasing influence of interconnectivity as the average interparticle distance is controllably reduced. We explore different regimes from those comprising low carrier densities, where recombination is limited by defect-assisted processes, relevant for photovoltaic applications, to those comprising large carrier densities where many-particle processes come into play, a scenario characteristic of light emitting devices. Results reveal that dot-to-dot transport and diffusion result in the exposure of the carriers to a larger defect landscape, which determines the efficiency of the different recombination processes occurring in the ensemble.

**RESULTS AND DISCUSSION**

**Structure and linear optical response of ligand-free perovskite QD networks**

Ligand free formamidinium lead bromide (FAPbBr$_3$) QDs were synthesized within nanoporous SiO$_2$ matrices by spin coating a solution of perovskite precursors in DMSO into it, following a procedure thoroughly described elsewhere.[23] Upon annealing at 100°C, perovskite QDs form in the voids of the SiO$_2$ matrix, as illustrated in **Fig. 1a**, where the precursor concentration $C_{prec}$ defines the overall amount of material that is injected into the host system. **Fig. 1b** shows a transmission electron microscopy (TEM) micrograph of an intermediate concentration, which illustrates the dispersion of crystallites attained in the matrix and their characteristic size distribution. The average crystal size, the number of particles formed per unit volume of the pores (or filling fraction, *ff*), and the average interparticle distance are affected by $C_{prec}$ (see **Table 1**), allowing to tune the interconnectivity without affecting their chemical composition, morphology or surface properties. All samples show clear signatures of quantum confinement effects, as evidenced by the shift towards higher energies of both the absorption edge (**Fig.1c**) and emission spectrum (**Fig.1d**) as the average QD radius, $\langle r \rangle$, is reduced, [20] although contributions to this shift from the dielectric environment cannot be discarded. [24,25] A reduction in QD size also leads to an enhancement in the continuous wave photoluminescence quantum yield (cw-PLQY) (**Fig.1e**). While this is the expected trend for an increasing level of quantum confinement, the key factor determining emission yield will be dictated by the average interparticle distance, as discussed below. It should be noticed that the excitonic peak at the absorption edge, clearly evident for the bulk case, is absent for the QDs under study. Such behavior has been found in QD solids in the past and associated with the presence of inter-dot electronic coupling and/or size polydispersity. [26,27,] The key structural parameters, $\langle r \rangle$ and *ff*, are obtained by fitting the spectral position of the PL maximum to the Brus equation [28] and performing porosimetry measurements, respectively. The Brus equation, while bearing a number of uncertainties due to the assumptions made, (see section S1.1 in the SI for a full discussion) has been proven to yield a good estimate of $\langle r \rangle$ for similar samples.[29] **Table 1** shows the estimated average size, which lies always below the Bohr radius for this material (8nm).[30] Both $\langle r \rangle$ and *ff* values are used to estimate the average spacing between the embedded QDs, $d_{QD}$, as summarized in **Table 1** (see the S.I. for details on the calculations). Please note that, from the data shown in **Table 1**, it is clear that while changing $C_{prec}$ modifies the QD size, the main effect is on the filling fraction and therefore on the average separation between QDs, $d_{QD}$, thus allowing control on inter-dot connectivity.

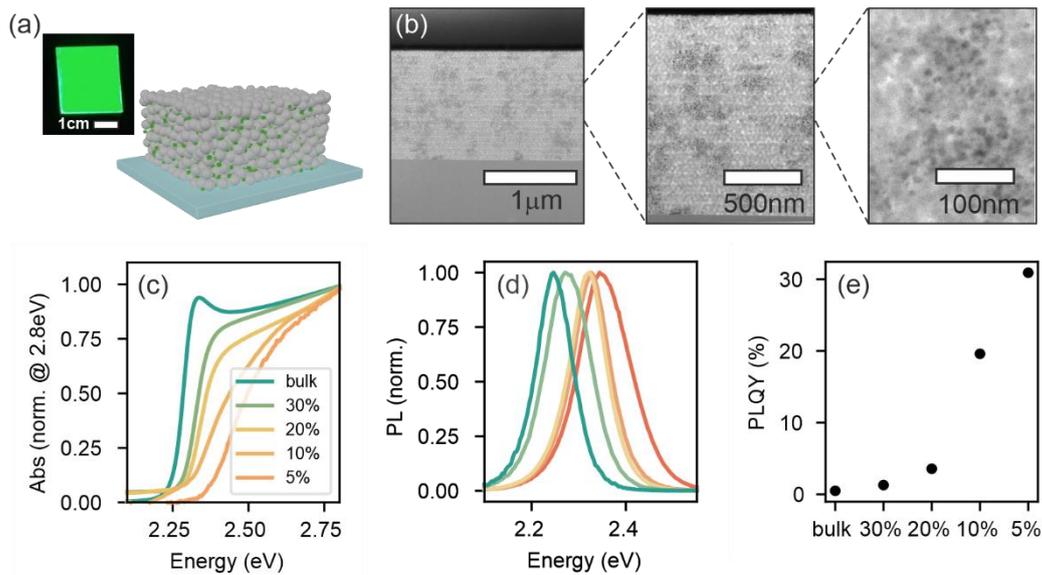

**Figure 1 Structural and linear optical characterization of ligand-free perovskite QDs embedded in porous matrices.** (a) Sketch of the analyzed materials system with an optical photograph of the thin film. (b) TEM cross section of $C_{prec} = 20\%$ at different magnifications. (c) Linear absorption normalized at 2.8eV. (d) Normalized PL of bulk and QD samples. The color code is the same as in (c). (e) PLQY of the different samples shown in (c) and (d).

**Table 1 Fill fraction, estimated average QD radius, and interparticle spacing extracted from porosimetry and optical measurements.** Please note that those values are approximated and are only meant to show the global trend between the individual samples and do not necessarily represent the exact values.

| $C_{prec}$ | $ff$ | $\langle r \rangle$ | $d_{QD}$ |
|---|---|---|---|
| 5 % | 0.0141 | 4.5 nm | 28 nm |
| 10 % | 0.0405 | 5.7 nm | 22 nm |
| 20 % | 0.092 | 6.3 nm | 15 nm |
| 30 % | 0.145 | 6.7 nm | 12 nm |

**Trap filling and carrier recombination dynamics in isolated and interconnected perovskite QD networks**

To understand the role of QD interconnectivity on the charge carrier recombination we start by studying the dynamics of the PL measured from the set of samples described above, which show a gradually varying interparticle distance $d_{QD}$. Results are presented in **Fig.2a-e**. The fluence conditions used in this experiment can be considered to be on the lower end, with an average of $10^{-5}$-$10^{-2}$ excitations per QD. For the case of a reference FAPbBr$_3$ bulk sample (**Fig.2a**) we obtain long PL lifetimes on the order of $\mu s$, which are strongly fluence dependent. The Shockley-Read-Hall model [31,32] can be applied to this scenario, where the recombination dynamics depend on the global population density of charge carriers that can freely move and distribute within the excited area, which is limited by the density of carriers trapped at defects. Here, the experimentally measured TRPL dynamics depend on the macroscopic density of electrons $n_e$, electron holes $n_h$ and trapped carriers $n_T$ and their recombination dynamics can be described by coupled rate equations (see Section S2.1 in the Supplementary Information), which account for the processes schematized in **Fig. 2f**. This model is well established for perovskite bulk samples [33,34,35] and yields a perfect fit of the TRPL dynamics of the bulk reference (gray solid lines in **Fig. 2a**), which provide defect densities and recombination rates in order with those reported in the literature (see fit values in Section S2.1 of the S.I.). This evidences that charge recombination under these conditions is dominated by trap filling and de-trapping dynamics. Interestingly, the fluence-dependence of the TRPL in the QD films becomes less pronounced as the filling fraction in the porous films decreases, and is eventually lost for the samples with the lowest amount of perovskite in the pore ($C_{prec} = 5\%$, i.e., those with the largest $d_{QD}$). In what follows, we analyze these dynamics in more detail.

For the cases of sparsely separated QDs ( $C_{prec} = 30 - 10\% \leftrightarrow d_{QD} = 12 - 22 nm$, **Fig. 2b-d**), the TRPL dynamics retain a fluence dependent behavior, albeit with markedly distinct trends with respect to the bulk sample. The PL decays now present two components: a fast (tens of ns) and intense initial decay followed by a much slower (several μs long) decay. As the fluence increases, the second component becomes more rapid and dominant in amplitude. Notably, the SRH model that has been successful in the bulk case fails to yield a good fit to these PL dynamics, indicating a more complex scenario than that depicted in **Fig. 2f**. For the extreme case of the largely separated QDs ($C_{prec} = 5\% \leftrightarrow d_{QD} = 28 nm$), the TRPL dynamics turn fluence independent (**Fig. 2e**). This can be interpreted in terms of isolated QDs where excitations are now well localized and dot-to-dot charge transport is not allowed. In this isolated QD scenario, the TRPL lifetime is determined by the ratio between non-radiative and radiative recombination pathways, characterized by rate constants $k_{nr}$ and $k_r$, respectively (**Fig. 2h**). For a homogenous distribution of monodisperse QDs, a single exponential decay is the expected characteristic of excitonic monomolecular recombination. [36] However, in our case (**Fig. 2e**) the TRPL presents a clear multi-exponential behavior, which is best fitted considering a lognormal distribution of decay rates. While this can be attributed to the QD size and shape polydispersity present in practically all collections of QDs, in our case the contribution from potentially different defect density or strain cannot be ruled out (see Section S2.2 and **Fig.S2.2** in the S.I.).

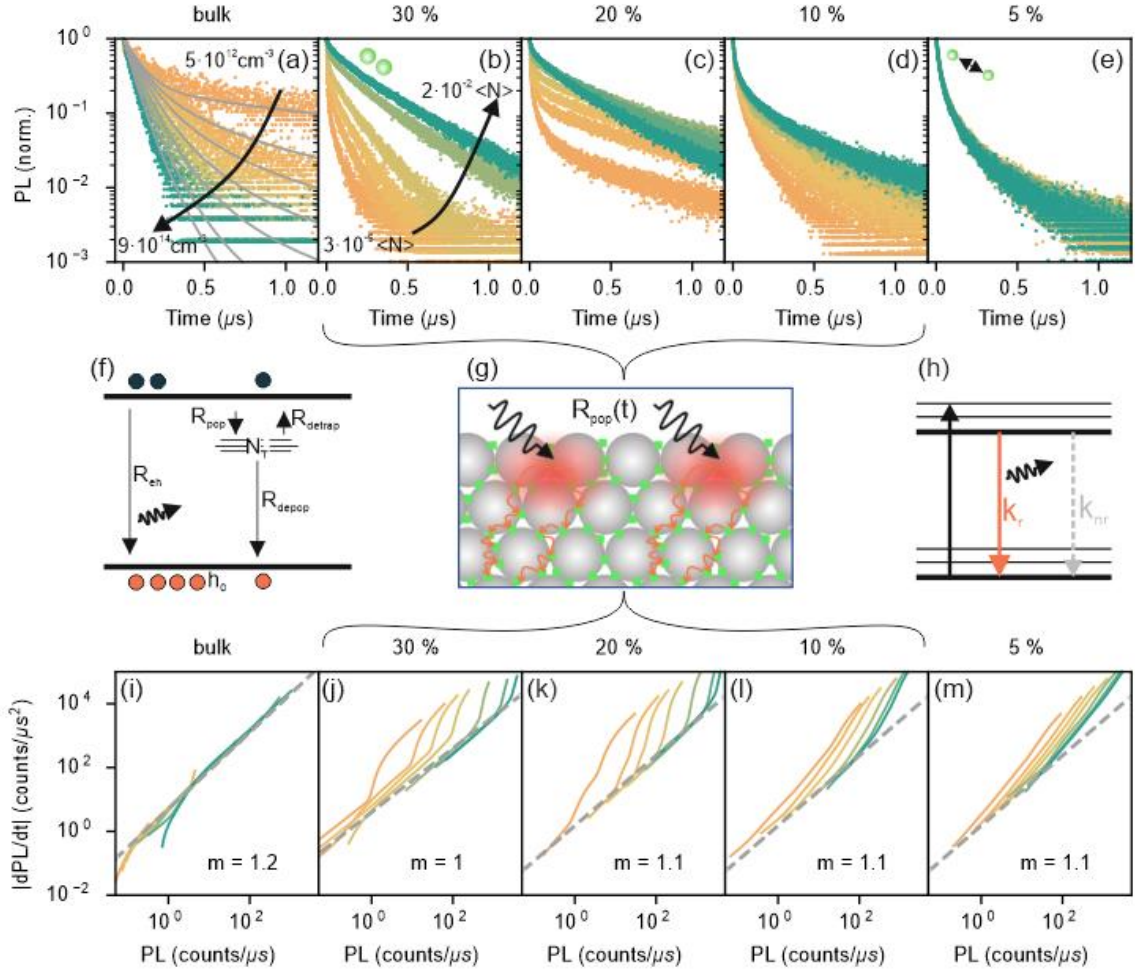

**Figure 2 TRPL measurements and corresponding charge carrier recombination models** (a)-(e) Fluence dependent TRPL measurements for bulk and QD samples with different $C_{prec}$. Maximum and minimum fluences are indicated as carrier densities (or average excitations per QD) for the bulk (or QD) cases. Gray solid lines in (a) indicate fits to the SRH model, while the gray line in (e) corresponds to the fit to a lognormal distribution of decay rates. Insets in (b) and (e) illustrate the expected crystal size and spacing (not to scale). (f) Model of the electron transition processes considered within the SRH model. (g) Drawing exemplifying the reduced diffusivity caused by the limited interconnection of QDs. (h) Model of an isolated emitter excitation and recombination. (i)-(m) Derivative of emitted photon flux against photon flux for the same samples in (a)-(e).

The effect of QD interconnectivity can be directly visualized and further analyzed by plotting the derivative of the emitted photon flux (PL in counts per second) as a function of the emitted photon flux. This derivative can be interpreted as proportional to an instantaneous effective carrier lifetime $\tau_{eff}(n)$, similar to the differential lifetime introduced by Kirchartz in Ref. 37, which bears information on the predominant recombination mechanism at a given time or carrier density (see below). Such a representation in the log-log scale is shown in **Fig. 2i-m** for all samples at all excitation densities employed. In order to interpret these graphs, it should be considered that, in a quasi-equilibrium condition and assuming the photophysical model

depicted in **Fig. 2f**, the first derivate of the photoluminescence can be written as (see Section S2.3):

$$\frac{dPL}{dt} = R_{eh}\big(R_{eh}n_e n_h \cdot (n_e + n_h) + R_{depop}n_T n_h \cdot n_e\big) \quad (1)$$

Considering that the emitted photon flux is proportional to the product of the free electron and hole densities ($n_e n_h$), equation (1) implies that the time variation of the photon flux is directly dependent on the total photon flux and hence the curves for all different excitation conditions must all line up in the log-log representation. Also, a monomolecular decay should result in a linear dependence of the $\frac{dPL}{dt}$ vs. PL curve with a slope close to 1, while a bimolecular one would yield a slope of around 1.5 (see discussion in SI Section S2.3 for more details). With this in mind, we can straightforwardly conclude that the PL decay dynamics of the bulk perovskite, shown in Fig 2i, is mainly dominated by trap-limited monomolecular recombination (slope m=1.1), with an effective lifetime, $\tau_{eff}(n)$, that depends only on the initial excitation density. On the other hand, a similar analysis of the curves plotted in Figs. 2j-l reveals that the emitted photon flux does not follow a monotonic trend with changing excitation density. The effective lifetime is now a time-dependent quantity, $\tau_{eff}(n,t)$, pointing at an out-of-equilibrium process that appears at early times in all measurements. This effect may be considered a direct consequence of the presence of interconnected QDs in the samples under analysis. Interconnectivity allows dot-to-dot charge transport and thus diffusion of carriers through the QD network to take place. So, when the excitation pulse gives rise to a high charge carrier population (as depicted in **Fig. 2g**), diffusion-mediated redistribution effects lead to a time-dependent reduction of the local average charge carrier density. At the same time, it allows carriers to sample a larger defect landscape, hence accounting for the reduced PLQY observed for these $C_{prec}$, as shown in **Fig. 1d**. These deviations correspond to the initial fast TRPL decay component observed in the TRPL of the connected networks (**Fig. 2b-d**), a signature similar to that reported for some bulk samples [38,39] and associated with diffusion and fast charge trapping. [40] In the most extreme case of isolated QDs (**Fig. 2m**) a set of parallel curves are observed, indicating a recombination dynamics that does not depend on the initial excitation density. In this case, all decays are multiexponential as faster (slower) emitters in an ensemble of QDs will contribute more at early (late) time scales.

**Interconnectivity Determines Emission efficiency in perovskite QD networks**

To further understand the impact of the interconnectivity on the macroscopic behavior of the perovskite QD networks under study, we next consider PLQY values as a function of fluence and temperature. This type of measurements allows identifying the prevalent recombination mechanism as a function of the injected carrier density, as well as the role played by defect trapping or lattice vibrations. We start the discussion with the bulk scenario where recombination dynamics depend on the charge carrier population density as described within the SRH model. By extending this model to higher-order recombination mechanisms such as Auger processes, and considering some assumptions,[41] the recombination dynamics can then be described in a simplified way within the ABC model,[35] where A, B, and C are nonradiative monomolecular, radiative bimolecular and nonradiative multiparticle recombination constants respectively. Assuming only the bimolecular processes to be radiative, the effective emission quantum yield can be calculated as:

$$PLQY(n) = \frac{Bn^2}{An+Bn^2+Cn^3}. \quad (2)$$

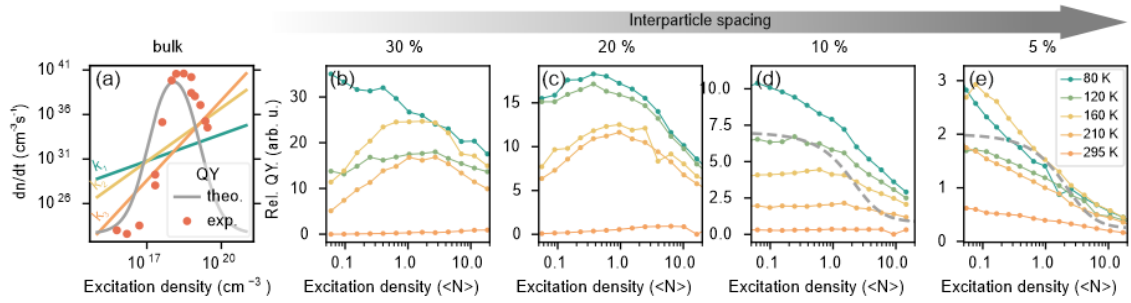

**Figure 3 Relative PLQY dependence on excitation density.** (a) Expected RT PLQY curve for a FAPbBr$_3$ bulk sample (gray line), where excitation density dependent values for A, B and C coefficients (green, yellow and orange lines) are extracted from TAS measurements. Red dots correspond to the experimental values extracted from a fluence dependent PL measurement. (b)-(e) Temperature and excitation density dependent relative PLQY measurements. Dashed gray lines are Poisson distributions indicating the onset of biexciton annihilation in a confined system.

The PLQY fluence dependence of the bulk material (red dots in **Fig.3a**) follows the theoretically expected bell shape (gray line), estimated considering the recombination coefficients extracted from the fitting of transient absorption spectroscopy (TAS) measurements using the ABC model. These experiments were carried out exploring recombination dynamics over a broad range of carrier densities (see section S3.1 and **Fig. S3.1** in the S.I.). The observed bell shape is indicative of increased dominance of radiative bimolecular recombination due to the filling of charge traps at increasing carrier population. [42] Above a certain threshold, non-radiative (trimolecular, Auger) recombination processes take over and the PLQY consequently decreases. Films with higher concentrations of QDs ($C_{prec} = 30\% \leftrightarrow d_{QD} = 12nm, C_{prec} = 20\% \leftrightarrow d_{QD} = 15nm$, Figs. 3b-c) in which inter-dot connection is more likely, show a similar PLQY bell-shaped curve (for the higher temperatures this may be more clearly seen in the normalized PLQY curves shown in **Fig. S3.2**). In these examples (see **Fig.3b-c**) it can be readily observed that, for very low excitation densities, $\langle N \rangle \ll 1$, trap state assisted recombination gives rise to a largely reduced PLQY, and that for $\langle N \rangle \gg 1$ non-radiative multiparticle processes are dominating, also causing a reduction of the PLQY. The optimum PLQY of interconnected QDs is expected at intermediate excitation densities, $1 < \langle N \rangle < 3-5$. As the temperature is reduced, two effects can be observed: an overall PL enhancement for each measured fluence and a relative increase in the emission for low fluences. The general PL enhancement is likely the combined result of lower non-radiative phonon-assisted recombination, as well as reduced inter-particle charge transfer and trapping. On the other hand, the higher relative rise in PL taking place for low fluences indicates a passivation of thermally activated defect states or a reduction in the trapping rate $k_{trap}$. In the regime where recombination is trap-assisted ($\langle N \rangle$ = 0.06, i.e., the ratio between the density of photons absorbed, calculated from the fluence employed and the absorptance, and the density of QDs in the film, estimated as explained in the Supplementary Information, section S1), an emission enhancement of almost 4 orders of magnitude is observed as the temperature is reduced, while emission dominated by multi-particle processes ($\langle N \rangle$ = 48) only rises by 1 order of magnitude (see **Fig. S3.3**), indicating that deactivation of trapping pathways is a key factor in the emission properties.

If we now consider the system with larger average separation between QDs ($C_{prec} = 5\% \leftrightarrow d_{QD} = 28nm$), where interconnectivity should be absent, a drastic change in the PLQY is observed (**Fig. 3e**). First, the overall PL enhancement as T is reduced is now much smaller (3-fold factor) than the one observed for the high $C_{prec}$ samples (3-4 orders of magnitude), very likely because temperature-induced changes in charge transport are now absent, given the larger inter-particle distance. The absence of dot-to-dot transport implies that the likelihood of a charge carrier getting trapped depends exclusively on the probability of a QD possessing a trap state, which is independent of the excitation density. Consequently, the monotonically growing initial part of the bell-shaped PLQY curve is lost. This effect also explains the much smaller PLQY observed for the films with shorter $d_{QD}$ at low fluences (see **Fig. 1b**). As the preparation methods of the different samples under analysis only differ in the concentration of precursors employed, we can assume that the defect density per QD is very similar in the $C_{prep}$= 30% and 5%, the potentially small variations between them not being enough to justify such large differences in PLQY. However, as spacing between QDs is larger, the defect landscape that carriers can access is reduced, hence increasing the chances of decaying through a radiative path. This finding can be generalized to any assembly of interconnected QDs: recombination in the low fluence regime is not only determined by the overall defect density in the material system but also by the presence of charge transport which allows carriers to sample a wider defect landscape. As for the high fluence regime ($\langle N \rangle$ >1 for highly confined systems), two-particle biexciton annihilation are responsible for the drop of PLQY observed.[43] Interestingly, the results attained for the $C_{prec} = 5\% \leftrightarrow d_{QD} = 28 nm$ and $C_{prec} = 10\% \leftrightarrow d_{QD} = 22 nm$ samples follows the Poisson function that provides the probability of exciting the same QD at a given average excitation density (dashed line in **Fig. 3d-e**), assuming any excitation above one leads to non-radiative biexciton annihilation.[44] In contrast, higher $C_{prec}$ samples show a multiparticle onset at higher excitation densities, which corresponds to Poisson distributions with a larger probability for radiative recombination allowed per QD, as evidenced in **Fig. S3.4**. This is consistent with a less efficient biexciton annihilation caused both by an increasing QD size (see **Table 1**) and most importantly, the loss of confinement induced by the transport of charges between neighboring dots, in good agreement with all previous observations.

**PL peak broadening in QD networks: Interplay of connectivity and nanocrystal size polydispersity.**

As stated in previous sections, the higher or lower level of connectivity between neighboring QDs affects the defect landscape to which carriers have access. Hence, connectivity becomes, along with size polydispersity and electron-phonon coupling,[45,46] a potential source of spectral broadening of the photoemission. While an exact estimation of the size distribution in our samples is not trivial, we expect the size dispersion to be very similar for all $C_{prec}$ employed. Thus, in order to estimate the role of connectivity in the emission of our QD networks, we performed low temperature measurements where electron-phonon coupling is largely reduced. With this in mind, we realized a comparative analysis of the PL emission spectra attained at 80K of the samples that display the largest difference in inter-dot separation, namely, $d_{QD}$=12 nm and $d_{QD}$=28 nm. Results, shown in **Fig.4**, evidence the reduction of electron-phonon coupling contribution, since strong variations in the spectral width of the PL, $w_{PL}$, are observed that are not present in RT measurements (see **Fig. S4.1**).

Further, strong differences in both $w_{PL}$ and its evolution with fluence are observed when QD connectivity is modified. If we consider the sample for which inter-dot separation is expected to be shortest ($C_{prec} = 30\% \leftrightarrow d_{QD} = 12 nm$) the PL is spectrally narrow and does not vary with excitation fluence (**Fig. 4a**). This behavior can be understood as the result of the funneling of photogenerated charges towards large QDs, which act as recombination centers and thus as the direct consequence of dot-to-dot transport. If we turn to the sample containing isolated QDs

($C_{prec} = 5\% \leftrightarrow d_{QD} = 28 nm$) a different behavior is observed (**Fig.4b**). On the one hand an overall spectrally broader PL is evident. This is to be expected as we are now measuring the additive PL from a collection of non-interacting QDs. On the other hand, a broader PL is measured as we increase the pump fluence. This fact can be explained as the consequence of the size dependent absorption cross section of the different QDs (see discussion in Section S4.1).

The hypothesis of inter-dot connectivity favoring carrier transfer is further supported by independent ultrafast TAS measurements (see **Fig. 4c,d**). The time evolution of the TAS spectra of the highly connected QD sample ($d_{QD}$=12 nm), presents a fast (1ps) initial narrowing of the ground state bleach (GSB) associated with the cooling of photoexcited hot-carriers in the sample (**Fig. 4c**). As the interparticle spacing is increased, the GSB feature becomes broader and presents a spectral narrowing (for energies around 2.5eV) taking place over a much longer times, up to 10 ps for the case of separated QDs ($d_{QD}$=28 nm) (**Fig. 4d**). The broader GSB feature is again the result of inhomogeneous broadening caused by the presence of a dispersion of isolated QDs and the longer times over which spectral narrowing takes place are similar to those reported in the past for cooling of hot-carriers in isolated lead halide perovskite colloidal nanocrystals.[47] A proper description of cooling dynamics in these systems certainly demands a more detailed study which is out of the scope of the present work.

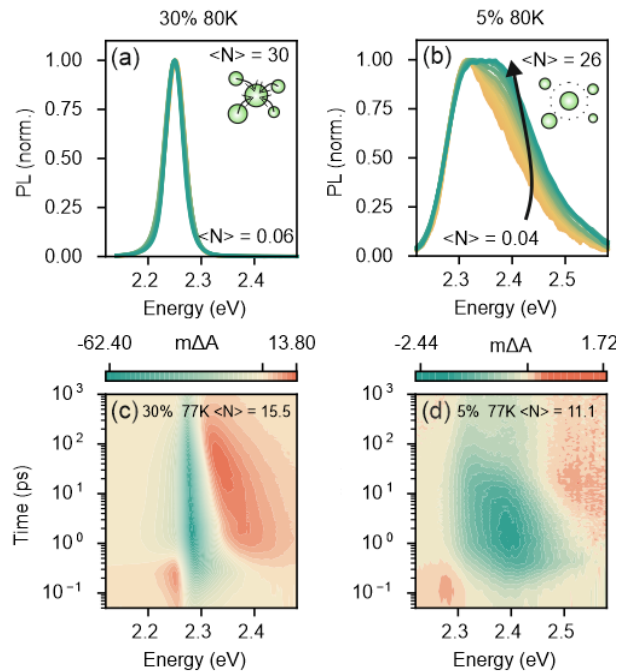

**Figure 4 Low temperature spectral shape of PL and TAS features.** (a)-(b): fluence dependent PL measurements at 80K for (a) 30% and (b) 5 % $C_{prec}$ sample. Insets illustrate the observed charge carrier funneling process. (c)-(d) High fluence TAS spectra represented in color maps for (c) 30% and (d) 5 % $C_{prec}$ samples.

**Multiparticle recombination effects in localized and interconnected emitters**

In this section we further explore the regime where a high carrier density is injected into the QD solid. This scenario, of relevance for lasing applications, is where non-radiative multiparticle processes can dominate recombination representing one of the main burdens for achieving optical gain.[3] To explore the role of QD connectivity in this regime we have performed fluence-dependent TAS measurements and studied the time dynamics of the GSB extracted as the minimum $\Delta A$ value at each time instant (**Fig. 5a,b**). In a bulk-like system, the above-described ABC model has been broadly employed to describe recombination as a function of the density of photogenerated carriers. [48] While good fits were obtained for the reference bulk sample presented before (see **Fig. S3.1**), deviations from the model become evident when we consider QD arrays. For the highly interconnected $d_{QD}$=12 nm sample, deviations of the experimental data from the model can be clearly seen when plotting the time derivative of the $\Delta A$ against $\Delta A$ (**Fig. 5c**). In analogy to the TRPL data shown above, this representation allows to discriminate between time-dependent ($\tau(n,t)$) and time-independent contributions ($\tau(n)$). Under equilibrium conditions, where charge carriers are homogenously distributed both spatially and energetically, the $\Delta A$ at the GSB is expected to be proportional to the charge carrier density. [49] In this scenario, a given charge carrier density should lead to a certain decay rate and consequently a fixed value for its time derivative. Changing the initial carrier density in a TAS experiment would lead to overlapping dynamics in a $\frac{d\Delta A}{dt}$ vs $\Delta A$ plot. For the $d_{QD}$=12 nm sample, we can observe early stage deviations which can be associated to charge carrier energy homogenization processes that comprise carrier diffusion as well as thermalization, the latter being strongly affected by confinement and Auger processes.[50] After ~2ns the TA lineshape does not show any significant change even at elevated excitation densities (compare **Fig. S4.2**) and the GSB decays overlap in their final stages, when carriers are assumed to be in equilibrium, and can be simulated with the ABC model as indicated by the gray line in **Fig. 5c**.

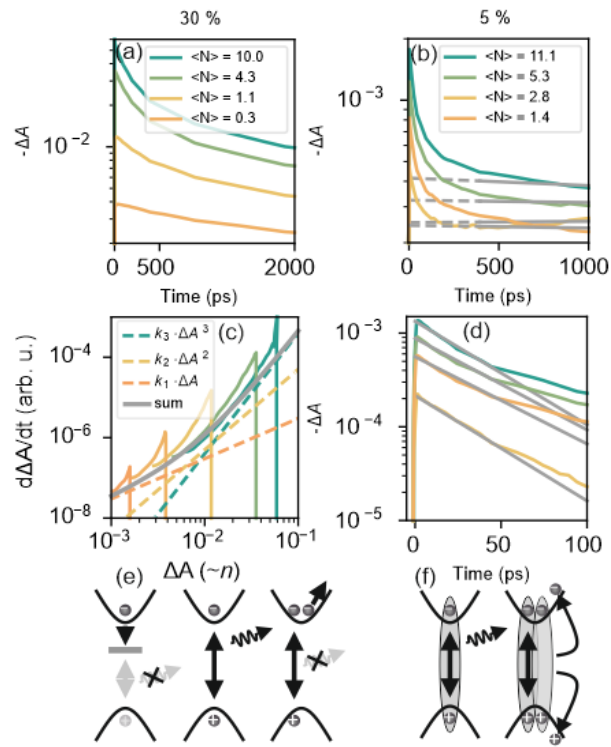

**Figure 5 Ground state bleach dynamics in samples with different degree of QD connectivity.** (a)-(b) GSB decays of (a) $C_{prec} = 30\%$ and (b) $C_{prec} = 5\%$ samples extracted as the minimum from the TAS signal. Gray dashed lines in (b) indicate ΔA background level mentioned in the text. (c) Derivative representation of the GSB data shown in (a). Gray line indicates sum of graphically approximated A, B, C decay rates (green, yellow and orange dashed lines respectively). (d) Difference of the GSB decay and the ΔA background indicated in (b). Gray lines represent single exponential fits (Fit values in SI). (e) Sketch of the proposed dominating ABC model for the $C_{prec} = 30\%$ sample. (f) Sketch of the biexciton annihilation process associated with the dynamics of the $C_{prec} = 5\%$ sample.

In contrast, small and isolated QDs, $d_{QD}$=12 nm, show a clearly distinct behavior, where the GSB decay can be separated into a fast initial component followed by slow decay into a $\Delta A$ background level independent of excitation fluence (**Fig. 5b**). This behavior is characteristic of quantum confined nanostructures presenting excitonic recombination where the fast-initial trace can be associated with the biexciton annihilation and the slower background with exciton recombination. [43,44] If the $\Delta A$ background level is subtracted from the general decay, the resulting signal can be fitted to a single exponential (**Fig. 5d**) which yields a fluence independent decay rate of 42 ps associated with the biexciton lifetime. This value is a factor of 2 smaller than previously reported for colloidal $FAPbBr_3$ NCs. [51] This small discrepancy is likely a combination of several factors such as the larger NC size used in Ref. 51 (10.5 nm) as well as the absence of any ligands in our samples.

**Conclusions**

In summary, we have unveiled the effect of dot interconnectivity on the charge carrier transport and recombination occurring in perovskite QD networks. In order to do so, we have taken advantage of a series of ligand-free QD assemblies with different degrees of connectivity (including isolated particles) attained by loading a nanoporous film with varying QD filling fractions. Employing different time-resolved spectroscopic techniques we have followed the carrier dynamics over a broad time and fluence range, which comprises carrier density regimes relevant for both photovoltaic as well as light-emitting applications. While samples presenting a low QD concentration can be described in the framework of isolated nanostructures, increasing the load leads to enhanced connectivity and a scenario where charge diffusion dictates the optoelectronic response. We find that charge diffusion, leading to a spatial homogenization within the QD solid, allows for charge transport and enhances the defect landscape accessible to each carrier. Our results imply that a judicious tuning of the connectivity among QDs should lead to a scenario where charge transport, needed for most optoelectronic applications, may be optimized by considering the balance between the demanded carrier density and the effect it has on the available recombination paths. Also, the analysis and concepts herein discussed may be extended and applied to QD solids of different morphology and composition, as they will allow to individuate the different recombination mechanisms for photoexcited carriers and pave the way for a precise device-oriented design.

**Methods**

*Sample fabrication*

*Materials:* Formamidinium bromide (FABr, GreatCell Solar Materials, 99,9%), Lead (II) bromide ($PbBr_2$, TCI, 99,99%), dimethylsulfoxide (DMSO, *Merck*, anhydrous 99.8%), methanol (MeOH, *VWR*, 98%), Poly(methylmethacrylate) (PMMA, *Alfa Aesar*, 99.9%) and chlorobenzene (CB, *Merck*, 99.9%) were used without additional purification steps.

*Preparation of $SiO_2$ nanoparticles porous scaffold:* Commercially available 30 nm $SiO_2$ nanoparticles (34% w/v in $H_2O$, LUDOX-TMA, Sigma-Aldrich) were diluted in methanol to 3% w/v to obtain a colloidal suspension. The diluted suspension was deposited on top of a low-fluorescence glass substrate via dip coating employing a withdrawal speed of 120 mm/min. To achieve a thickness of around 1 µm, the deposition was repeated 15 times and consequently annealed at 450ºC for 30min to remove any residual organic component in the matrix and to improve its mechanical stability.

*Synthesis of FAPbBr$_3$ QDs within nanoporous silica scaffold:* A perovskite solution precursor was prepared using FABr and PbBr$_2$ powders dissolved in DMSO in a 1:1 molar ratio at different concentrations. By spin-coating (5000 rpm for 60 s) the solution was infiltrated in the void spaces of the scaffold, followed by a heat treatment at 100ºC for 1hour to obtain FAPbBr$_3$ QDs within the pores of the matrix. The synthesis was conducted in a protected N$_2$ environment of a glovebox.

*Electron Microscopy:* TEM images were acquired using a FEI Talos F200S scanning/transmission electron microscope operated at 200 kV from a lamellae obtained with a focused ion beam (FIB, Carl Zeiss, Auriga).

### Optical characterization

*Absolute Photoluminescence quantum yield measurements*: PLQY was measured in a commercial fluorometer (*Edinburgh* FLS1000) using an excitation wavelength of λ = 450nm with the sample contained in an integrating sphere accessory.

*Time resolved photoluminescence measurements:* Time resolved measurements were carried out using a 200kHz pulsed white light source (*NKT Fianium SuperK*) filtered by an acousto-optic tunable filter to an excitation bandwidth ranging from 450-470nm in combination with an *id100* single photon detector (*ID Quantique*) and a *SPC-130-EMN* time correlated single photon counting detection card (*Becker and Hickl*). For precise fluence estimation, the spot size was controlled with an *Ophir SP932U* beam profiling camera and power were continuously measured with a beam sampling configuration.

*Relative Photoluminescence quantum yield measurements*: Temperature dependent relative PLQY measurements where performed using a home-built setup employing a pulsed (1kHz) ultrafast laser (λ$_c$=1030nm) *Pharos* coupled to an *Orpheus* collinear optical parametric amplifier (both from *Light conversion*) set to emit ~200 fs, 450nm excitation pulses. The power and beam profile were continuously measured with a beam sampling configuration and adapted with automated filter wheels. The sample was placed in an *OptistatDN* cryostat (*Oxford instruments*) and detection is carried out with a *Flame* spectrometer (*Ocean Insight*), where the acquisition time was adapted to increase the dynamic detection range and detection counts where normalized to the acquisition time. All hardware components and data acquisition where fully automated and controlled by a self-written *Python* program.

*Transient absorption spectroscopy measurements:* Transient absorption measurements where performed using a commercial transient absorption spectroscopy setup (*HELIOS* Femtosecond Transient Absorption Spectrometer from *Ultrafast Systems*) using the same excitation as with the relative photoluminescence QY measurements. Cryogenic measurements where performed employing the above-mentioned cryostat.

**Acknowledgements:**

HM is thankful for the financial support of the Spanish Ministry of Science and Innovation under grant PID2020-116593RB-I00, funded by MCIN/AEI/10.13039/501100011033, and of the Junta de Andalucía under grant P18-RT-2291 (FEDER/UE). HM, JFGL, and DOT acknowledge financial support from the European Union's Horizon 2020 research and innovation program under the Marie Skłodowska-Curie grant agreement No 956270 (Persephone). ARSK acknowledges the start-up funds provided by Wake Forest University and funding from the Center for Functional Materials and the Office of Research and Sponsored Programs at WFU.

# Supporting Information for:

# Effect of connectivity on the carrier transport and recombination dynamics of perovskite quantum dot networks

David O. Tiede[1], Carlos Romero-Pérez[1], Katherine A. Koch[2], K. Burak Ucer[2], Mauricio E. Calvo[1], Ajay Ram Srimath Kandada[2], Juan F. Galisteo-López[1]*, Hernán Míguez[1]*.

[1]*Instituto de Ciencias de Materiales de Sevilla (Consejo Superior de Investigaciones Científicas-Universidad de Sevilla), C/Américo Vespucio, 49, Sevilla, 41092, Spain.*

[2] *Department of Physics and Center for Functional Materials, Wake Forest University, 1834 Wake Forest Road, Winston-Salem, NC 27109, USA.*

*E-mail: juan.galisteo@csic.es, h.miguez@csic.es

## Contents

# S1 Structural characterization
## S1. 1 Size and interparticle distance estimation from fill fraction and optical measurements

Average perovskite quantum dot (QD) sizes for the different samples under study were estimated by using the Brus equation, [1] an approach that has proven to yield a good estimate in the past for similar samples: [2]

$$E_{g,nano} = E_{g,bulk} + \frac{h^2}{8\mu R^2} - \frac{1.786 e^2}{4\pi\varepsilon_0\varepsilon_r R} \qquad (1)$$

where the values for effective mass (μ) and dielectric constant (ε_r) were taken from reference [3] and the bandgap estimated from PL measurements of the bulk sample. This equation assumes a fixed set of parameters for effective mass of electron $m_e^*$, hole $m_h^*$ as well as for the dielectric constant and assumes this set to remain unchanged under confinement. Further, the estimation of the bulk bandgap from the emission energy contains uncertainties since the effect of an excitonic transition is not considered. In addition, possible effects from size distribution and charge funneling effects and modifications in the electronic band structure through e.g. strain are not reflected in eq. 1. Despite the uncertainties introduced by the mentioned effects, we employ the Brus equation to get an estimation of particle sizes to obtain a figure of comparison between the different samples rather than to obtain an exact value for the actual size and size distribution. The estimated average sizes are shown in **Table S1**.

**Table S1 Fill fraction, estimated average QD radius and interparticle spacing** extracted from porosimetry and optical measurements.

| Precursor concentration $C_{prec}$ | Fill fraction $ff$ | Estimated <r> | # of QDs per m³ | $d_{center}$ | $d_{interp}$ |
|---|---|---|---|---|---|
| 5 % | 0.0141 | 4.5 nm | $3.7 \cdot 10^{22} m^{-3}$ | 37 nm | 28 nm |
| 10 % | 0.0405 | 5.7 nm | $5.2 \cdot 10^{22} m^{-3}$ | 33 nm | 22 nm |
| 20 % | 0.092 | 6.26 nm | $9.0 \cdot 10^{22} m^{-3}$ | 28 nm | 15 nm |
| 30 % | 0.145 | 6.67 nm | $8.0 \cdot 10^{23} m^{-3}$ | 25 nm | 12 nm |

To estimate the interparticle spacing between QDs, we assume the particles to be equally spaced within the host volume. In that case, in a volume V filled with N particles, the average volume v per particle is given by

$$v = \frac{V}{N} \qquad (2)$$

Where, assuming spherical particles, the volume can be expressed as

$$v = \frac{4}{3}\pi r^3 = \frac{V}{N} \qquad (3)$$

To obtain the average distance between particle centers, the previous equation can be rewritten as:

$$d_{center} = 2r = 2\sqrt[3]{\frac{3V}{4\pi N}} \qquad (4)$$

The average void distance between particles can consequently be estimated as

$$d_{interp} = d - 2<r> \qquad (5)$$

From this estimation, the values shown in **Table 1** can be extracted. It becomes evident that the average particle size is only changing by a factor of 1.5, while the interparticle spacing is more than doubled.

## S2 Trap filling dynamics in isolated and interacting QD networks

### S2.1 Shockley – Read – Hall charge carrier recombination model.

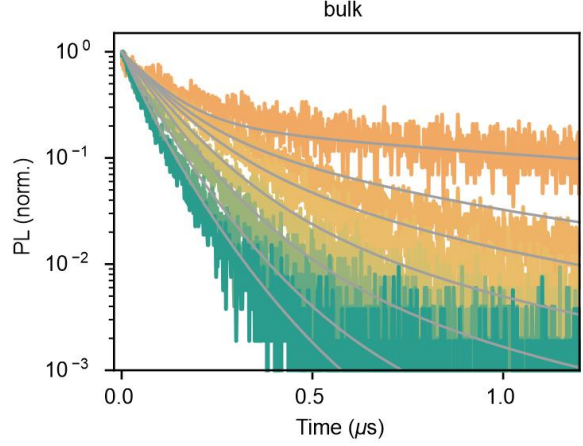

**Figure S2.1 Shockley- Read-Hall model for bulk FAPbBr$_3$.** Normalized fluence dependent TRPL data with fluences ranging from $5.0 \cdot 10^{12} cm^{-3}$ (orange curve) to $8.7 \cdot 10^{14} cm^{-3}$ (green curve). Grey lines represent global fit to the SRH model described below.

The Shockley -Read – Hall (SRH) model describes the charge carrier recombination dynamics with a set of differential equations for the change in the population density of electrons $n_e$, trapped electrons $n_T$ and holes $n_h$.[4] Given the low excitation regime, we exclude Auger effects from the model and assume the following rate equations:

$$\frac{dn_e}{dt} = -R_{pop}(N_T - n_T)n_e + R_{detrap}n_T - R_{eh}n_e n_h \quad (6)$$

$$\frac{dn_T}{dt} = +R_{pop}(N_T - n_T)n_e - R_{detrap}n_T - R_{depop}n_T n_h \quad (7)$$

$$\frac{dn_h}{dt} = -R_{depop}n_T n_h - R_{eh}n_e n_h \quad (8)$$

, where $R_{pop}$, $R_{detrap}$, $R_{eh}$, $R_{depop}$ describe the electron trapping rate, the detrapping rate of an electron returning to the conduction band, the radiative electron-hole recombination rate and the non-radiative trapped electron-hole depopulation rate, respectively. Note that we do not discriminate between possible electron or hole defects and that the same result could be reproduced with hole traps instead of electron traps. The global fit of the experimental time-resolved photoluminescence (TRPL) data results in the following rate constants:

$$R_{pop} = 1.95 \cdot 10^{-14} \text{ m}^3\text{s}^{-1}$$

$$R_{detrap} = 2.03 \cdot 10^6 \text{ s}^{-1}$$

$$R_{depop} = 1 \cdot 10^{-13} \text{m}^3\text{s}^{-1}$$

$$R_{eh} = 3.61 \cdot 10^{-15} m^3 s^{-1}$$

$$N_T = 4.22 \cdot 10^{20} m^{-3}$$

## S2.2 Distribution of effective lifetimes in isolated QDs

Isolated QD samples show a distinct behavior as compared to interconnected ones both in static and dynamic PL measurements. The PL spectrum for the $C_{prec}$=5% sample (**Fig. S2.2a**) is broader than the rest of the samples under consideration where a certain degree of interconnection is expected. Further, TRPL measurements show a spectral dependence, which is expected for a sample comprising a distribution of independent QDs. The PL decay when only the high energy components of the PL spectrum are considered (green band in **Fig. S2.2a**) is much faster than for the low energy components (orange band), which match the overall integrated decay (**Fig. S2.2b**). The PL decays show a multi-exponential behavior that could be best fitted to a lognormal distribution of decay rates (**Fig. S2.2c-e**).[5] From these fits a lifetime distribution can be retrieved for each spectral range (**Fig. S2.2f-h**) where it becomes evident the presence of a distribution of isolated QDs with different emitting properties. Here it should be mentioned that the presence of QDs with faster PL decay and higher energetic emission cannot be exclusively associated with the QD size but other factors must be considered such as a varying density of defects, modifications in the energy levels due to a modified crystal geometry or the presence of strain.

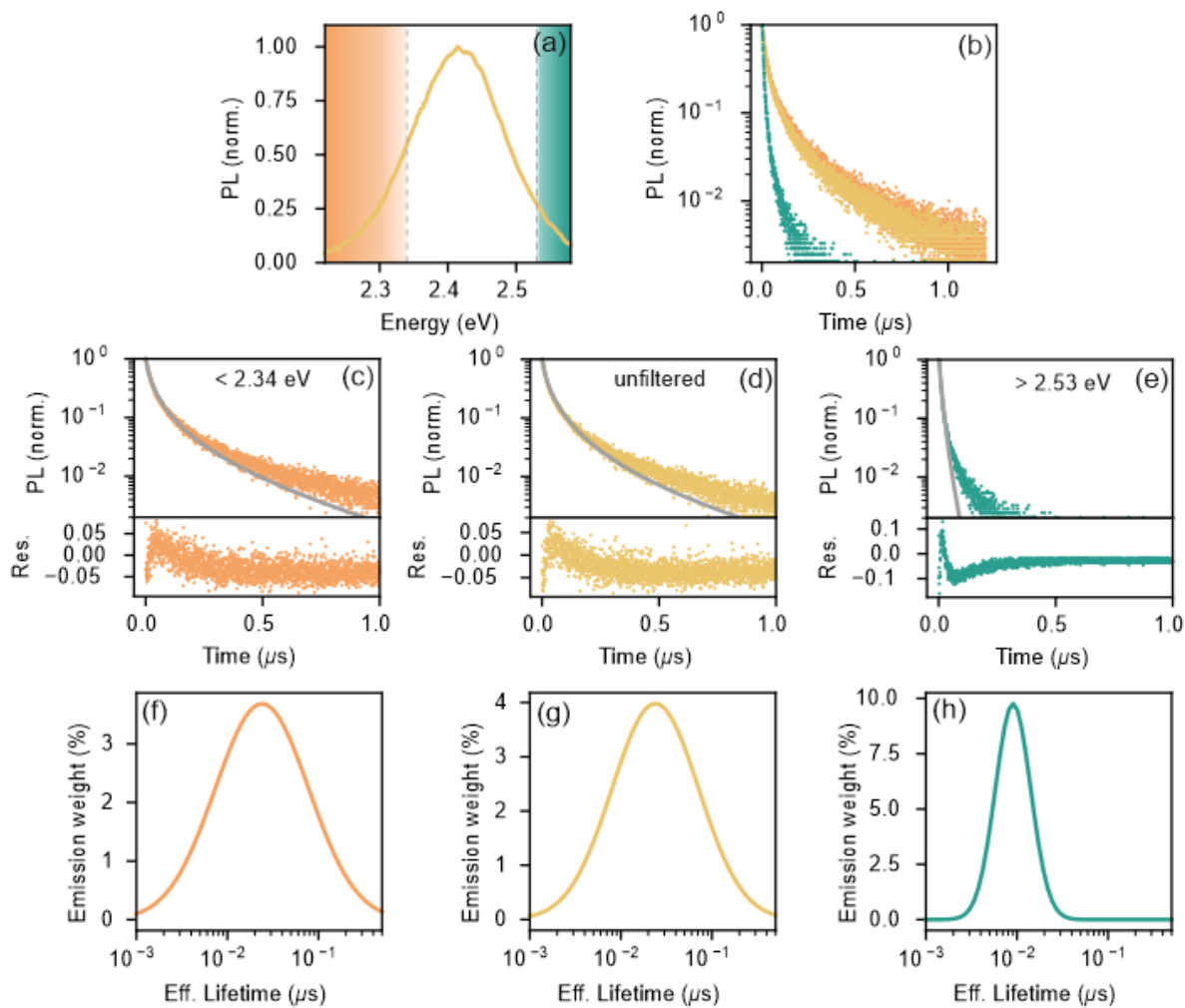

**Figure S2.2 Spectrally resolved effective lifetime distributions of isolated QDs ($C_{prec} = 5\%$).** (a) PL emission spectrum of $C_{prec} = 5\%$ at RT. Shaded green (orange) area indicates high (low) energetic emission ranges considered in the following panels. (b) Comparison of TRPL decays detected without spectral detection filter (yellow dots), detected with energies below 2.34eV (orange dots) and above 2.53eV (green dots). Color code used for the different detection ranges is identical for all panels. (c)-(e) Fit (grey lines) of TRPL decays to a log

normal distribution of effective lifetimes. Bottom panel shows residual values of fit. (f)-(h)Emission weight of the effective lifetimes extracted from (c)-(e).

The mean effective lifetime of the high energetic emission is ~ 10 ns= $1 \cdot 10^{-8} s$. This corresponds to the fastest possible emitter, where no emission delaying through defect trapping and detrapping or other delaying effects such as polaron formation or modification of the electronic band structure through strain take place. In the SRH model, the direct electron-hole recombination $R_{eh} = 3.6 \cdot 10^{-15} cm^3 s^{-1}$ can be used to estimate the radiative recombination rate in a defect free QD, assuming $V_{NC} = 3.8 \cdot 10^{-25} m^3$, two charges per excitation and $\tau^{-1} = k\rho = R_{eh} \cdot \frac{2}{V_{Nc}} = (5.3 \cdot 10^{-11} s)^{-1}$ ($1.6 \cdot 10^{-10} s$ for $C_{prec} = 30\%$. This effective lifetime extracted from the population density dependent SRH model can be further slowed down through retardation effects such as thermalization, exciton formation, polaron formation or (surface-) strain induced deformation of the electronic band structure, such that the mismatch between the SRH value and the value extracted from the log-normal distribution can be explained. Nevertheless, this simplified calculation demonstrates that pristine QD are likely to show a strongly accelerated recombination dynamic and that a reduction in $V_{NC}$ leads to an acceleration in recombination time.

## S2.3 Photon flux representation of TRPL data

TRPL data from the different samples under consideration (shown if **Fig.2a-e** of the main text) were also represented as photon flux vs. its time derivative in order to identify out-of-equilibrium processes taking place during charge carrier recombination. In order to perform such representation (**Fig.2i-m**) TRPL data were approximated with a multiexponential decay function, where the number of exponential decay terms was varied in order to reproduce the experimental data as close as possible. The multiexponential decay function was derived analytically and plotted against the amplitude of the PL flux for each data point.

Charge carrier recombination models mentioned in the main text (SRH and ABC models) describe the recombination of excited carriers in an equilibrium situation where they are homogenously distributed in space within a semiconductor and lying at the maximum (minimum) of the valence (conduction) band. Under this assumption, the time derivative of the photon flux should follow a monotonous trend where the slope will vary reflecting which process dominates recombination (monomolecular, bimolecular or higher order processes) for a given carrier density. If different measurements are performed for varying initial injected carrier densities $n_0$, experimental data should fall along the mentioned monotonous trend.

Any out-of-equilibrium process will then represent a deviation from this expected trend. These situations can imply energetic deviations (carriers are not lying at band extrema as in the case of a population of hot carriers) or carriers are undergoing spatial diffusion before reaching a homogeneous spatial distribution within the matrix. With the time resolution of our experiments we are probing a population of thermalized carriers thus any deviation from the expected trend should come from carrier diffusion processes. Under these circumstances, when plotting the data for several measurements with varying $n_0$, an out-of-equilibrium situation should appear in the initial stages where a spatially homogeneous distribution of carriers has not been achieved. Such deviations are absent in the bulk and non-connected QD ($C_{prec}$=5%) samples indicating that charge diffusion is either too fast (bulk) or not present (separated QDs). For intermediate samples an out of equilibrium process becomes evident at the beginning of each measurement showing the presence of diffusion. The reason why diffusion is taking such long times as to become evident in this representation is due to two (connected factors): charges must be transferred from one QD to the adjacent one and in this process, they have access to a larger defect landscape (as they are able to probe traps states present at neighboring QDs).

The longer, converged tails of the decay curves can be described within an equilibrium situation, where trap filling ($R_{pop}(N_T - n_T)n_e$) and detrapping ($R_{detrap}n_T$) have equilibrated. In this scenario, eq.6 and eq.7 simplify to:

$$\frac{dn_e}{dt} = -R_{eh}n_e n_h \quad (9)$$

$$\frac{dn_h}{dt} = -R_{depop}n_T n_h \quad (10)$$

Further, the trapped carrier density can be expressed as a function of electron density as:

$$\frac{dn_T}{dt} = +R_{pop}(N_T - n_T)n_e - R_{detrap}n_T - \underbrace{R_{depop}n_T n_h}_{\ll R_{pop}} \quad (11)$$

$$n_T = \frac{R_{pop}N_T n_e - \frac{dn_T}{dt}}{(R_{detrap} + R_{depop}n_h + R_{pop}n_e)} \approx \frac{R_{pop}N_T n_e}{(R_{detrap} + R_{pop}n_e)} \quad (12)$$

, as $R_{depop}n_T n_h, \frac{dn_T}{dt} \ll R_{pop}N_T n_e$. Next, due to charge neutrality, the hole density can be expressed as a function of electron and trapped electron density:

$$n_h = n_e + n_T = n_e + \frac{R_{pop}N_T n_e}{(R_{detrap} + R_{pop}n_e)} \quad (13)$$

The y-axis in the photon flux representation is given by the derivate of the PL signal, which can be now written as:

$$\frac{dPL}{dt} = R_{eh}\frac{dn_e n_h}{dt} = R_{eh}\frac{dn_e}{dt}n_h + R_{eh}n_e\frac{dn_h}{dt}$$

$$= R_{eh}\big(R_{eh}n_e n_h \cdot (n_e + n_h) + R_{depop}n_T n_h \cdot n_e\big) \quad (14)$$

The x-axis of the photon flux representation is given by the amplitude of the TRPL signal, which is proportional to the densities of electrons and holes ($PL \propto R_{eh}n_e n_h \propto n_e n_h$). Now, two scenarios can be considered:

- Trap recombination dominated, monomolecular regime ($n_h, n_T \gg n_e$):
  Here, the majority of electrons get trapped, such that $n_T \approx n_h$. Since $R_{depop} > R_{eh}$, **eq. 14** simplifies to:

  $$\frac{dPL}{dt} \propto R_{depop}n_T n_h n_e \quad (15)$$

  In this regime, the recombination dynamics are effectively dominated by two charge carrier densities, such that a representation of the product of two charge carrier densities against the product of two charge carrier densities results in a slope of $m = 1$ in a log-log representation (**see Fig. S2.3**).

- Electron-hole recombination dominated, bimolecular regime ($n_e \approx n_h$):
  Here, the number of free electrons exceeds the number of trapped electrons ($n_e > n_T$). In this case, in **eq. 14** the electron-hole recombination dominates, such that

  $$\frac{dPL}{dt} \propto R_{eh}n_e n_h \cdot (n_e + n_h) \quad (16)$$

In this regime, the recombination dynamics are dominated by the product of three charge carrier densities. A representation of the product of three charge carrier densities against the product of two charge carrier densities results in a slope of m =1.5 in a log-log representation (**see Fig. S2.3**).

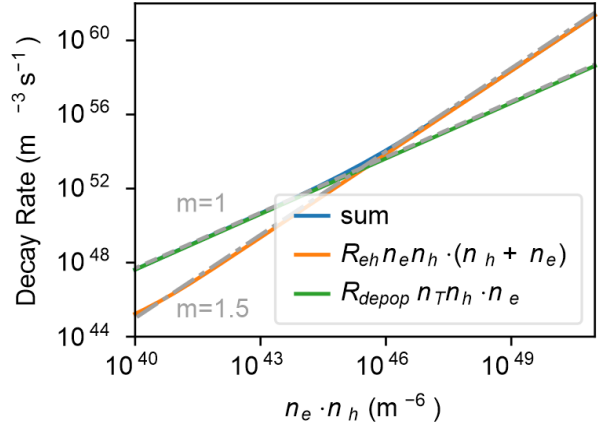

**Figure S2.3 Simulation of the decay rates of electron-hole and trap depopulation contributions.** At low charge carrier densities, the trap depopulation mechanism with a slope of $m = 1$ dominates. The higher charge carrier density regime is dominated by the electron hole recombination that has a slope of $m = 1.5$. The curves are simulated with the fit values of section S2.1.

# S3 Temperature and fluence dependent PLQY data
## S3.1 ABC model for bulk FAPbBr$_3$

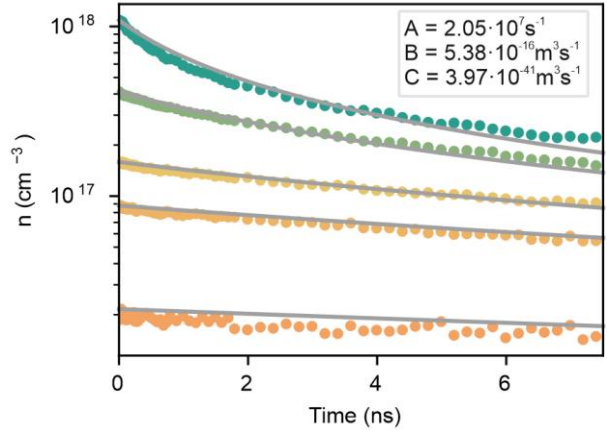

**Figure S3.1 GSB decay dynamics of FAPbBr3 bulk.** Fluence dependent GSB dynamics fitted globally to the ABC model as described in the main text.

Room temperature fluence dependent transient absorption spectroscopy (TAS) was performed on FAPbBr$_3$ bulk samples and the GSB taken as the minimum value of the TAS spectra. The $\Delta A$ GSB signal was converted into charge carrier density by establishing a relationship between the injected charge carrier density and initial GSB signal for each fluence. The time evolution of the GSB was adjusted to the ABC model:

$$\frac{dn}{dt} = -A \cdot n - B \cdot n^2 - C \cdot n^3 \qquad (17)$$

Excellent fits were obtained (see **Fig. S3.1**) from which values for the three recombination rates were obtained: $A = 2.05 \cdot 10^7 s^{-1}$, $B = 5.38 \cdot 10^{-16} m^3 s^{-1}$ and $C = 3.97 \cdot 10^{-41} m^6 s^{-1}$.

## S3.2 Normalized excitation density and temperature dependent PLQY

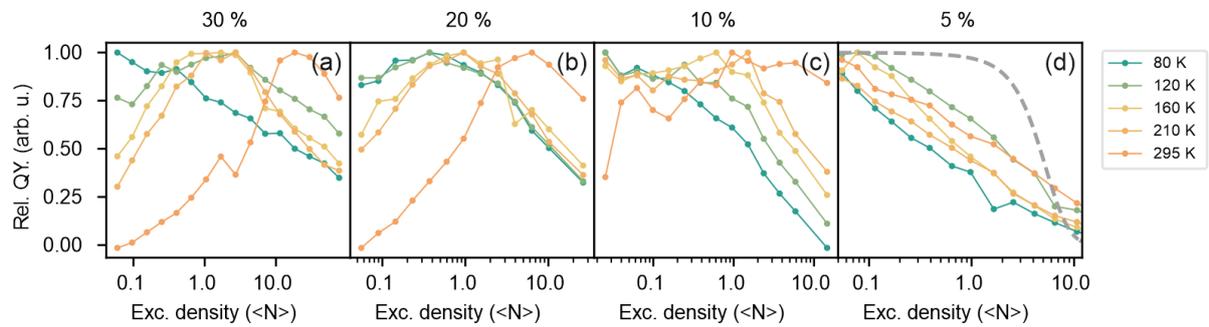

**Figure S3.2 Normalized PLQY data for different QD samples.** (a) – (d) PLQY curves normalized to their maximum PLQY for different $C_{prec}$ values. Dotted line in (d) indicates normalized PLQY expected for a Poisson distribution of excitations per QD where only one recombination per QD is radiative.

## S3.3 Temperature dependent PL enhancement

As discussed in the main text, the temperature dependence of the PL in the NC solids under study is strongly influenced by the carrier density which determines which recombination regime dominates. Here we discuss two extreme cases. For the highly interconnected samples with $C_{prec} = 30\%$ (**Figure S3.3a**) emission undergoes a drastic enhancement of 4 orders of magnitude as T is reduced from RT to 77K in the low fluence scenario, where charge carrier recombination is dominated by trap-assisted processes. This is, as discussed, likely a combination of passivation of thermally activated defect states and a reduction in the trapping rate $k_{trap}$. For higher fluences the PL enhancement is only slightly above an order of magnitude.

For the isolated case ($C_{prec} = 5\%$) (**Figure S3.3b**), enhancement across all fluences is much weaker (below an order of magnitude) since the role of defect assisted recombination is reduced as carriers cannot access defect traps in neighboring QDs. In addition, this sample shows characteristic PLQY enhancement around phase transition temperatures. [6] Such enhancement becomes more evident for isolated QDs as charge funneling processes, that can hide out local variations induced by the phase transition of individual QDs, are prevented. A detailed discussion of those dynamics is beyond the scope of this work.

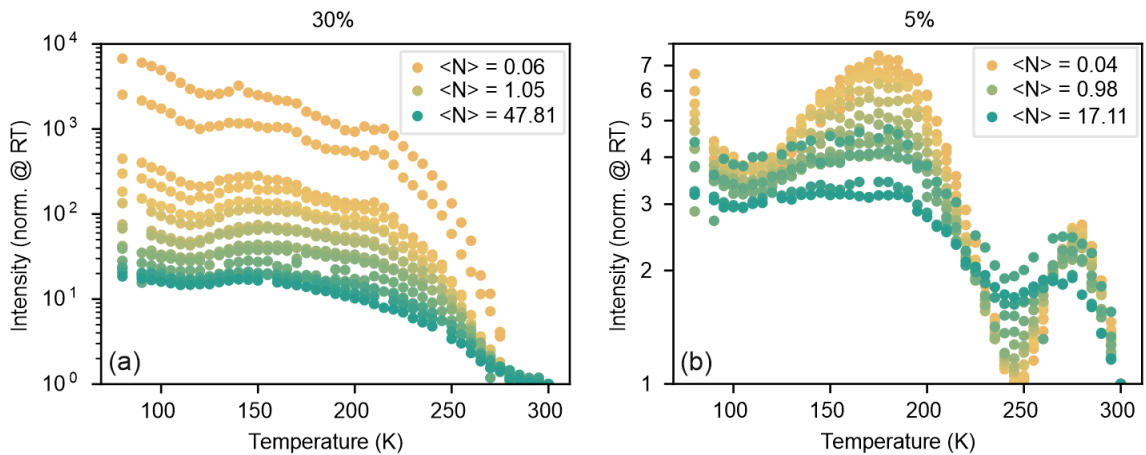

**Figure S3.3 PLQY values for different fluences normalized at room temperature.** Relative PLQY values for different fluences for (a) $C_{prec} = 30\%$ and (b) $C_{prec} = 5\%$ normalized at the RT value.

## S3.4 Simulation of Poisson distributed excitation scenarios

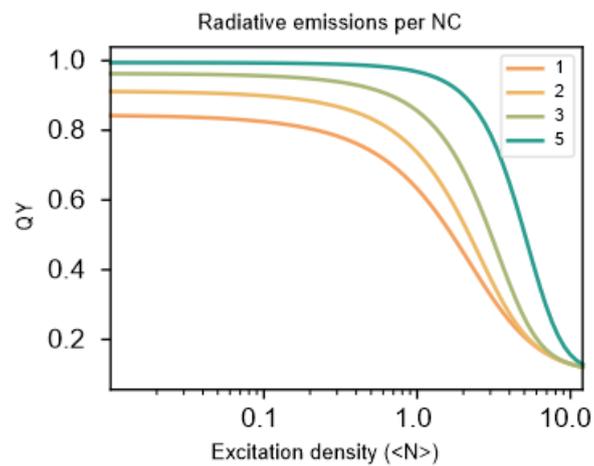

**Figure S3.4 Simulated PLQY for Poisson distributed excitation densities.** PLQY curves for different threshold levels of radiative recombinations per QD per excitation pulse ranging from 1 to 5 radiative emission per QD.

The probability of an individual QD to absorb an incoming photon can be described with a Poisson distribution. [7] The fluence onset where the PLQY starts to decay depends on the number of permitted radiative emission processes $Recomb._{rad}$ per excitation pulse in each QD. If the number of absorbed photons is higher than $Recomb._{rad}$, the PLQY starts to decrease as simulated in **Fig S3.4**.

# S4 Size distribution and charge carrier funneling effects

## S4.1 Spectral analysis at different temperatures and fluences

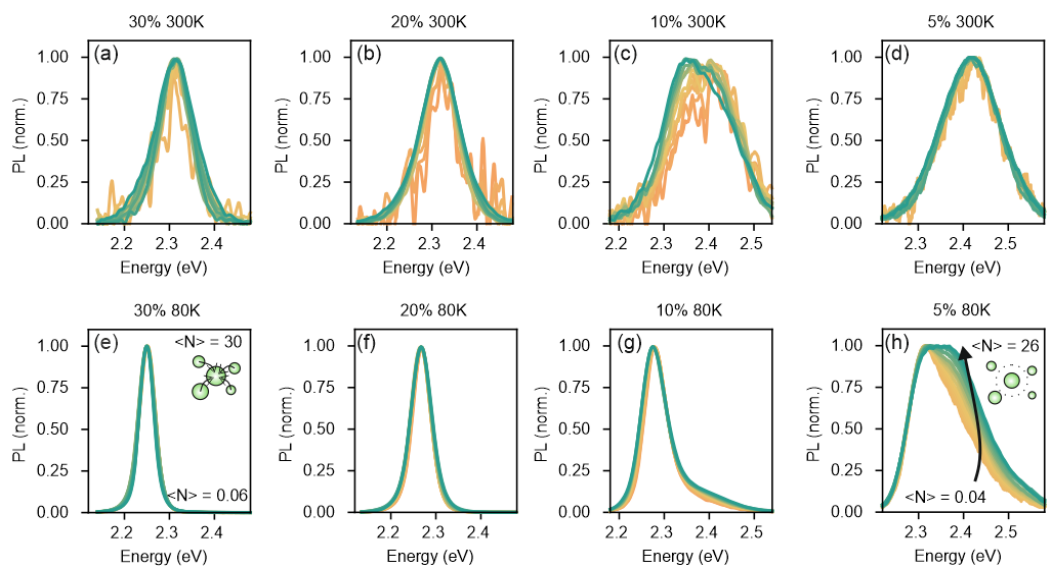

**Figure S4.1 Emission spectra of samples with different $C_{prec}$.** (a)-(d) RT normalized emission spectra. A gradual increase in peak width can be observed from 30% to 5 %. (e)-(h) Normalized emission spectra at low temperature. All graphs show excitation densities ranging from 0.04 (orange) to 30 <N> (green). A strong increase in FWHM can be observed in the 5% case. Insets show the expected charge funneling scenario in each case.

The emission bandwidth of interconnected ($C_{prec} = 30\%$) is narrower than the the isolated ($C_{prec} = 5\%$) case. This observation becomes more evident when considering emission spectra at 80K, where thermal broadening is reduced. In the interconnected case, the narrow spectral width of the PL can be rationalized considering a charge funneling process, where emission occurs from the energetically lowest site of the system. In contrast, in the isolated case, the emission occurs also from higher energetic states, as a funneling to the lowest state is prohibited. In addition, at 80K a fluence dependence of the emission spectrum becomes evident in the $C_{prec} = 5\%$ sample. This is interpreted as a higher absorption probability of larger QDs with lower emission energy, as their spatial absorption cross section is larger than the spatial absorption cross section of smaller ones. This higher absorption probability leads to an enhanced emission weight of larger QDs at low fluences. As fluence increases, the radiative emission of large QDs saturates with excess charge carriers recombining non-radiatively, whereas smaller ones continue to receive a larger load of excitations that are emitted radiatively. This mechanism further evidences the isolated character of the $C_{prec} = 5\%$ sample.

## S4.2 Energy resolved state filling dynamics evidenced by TAS

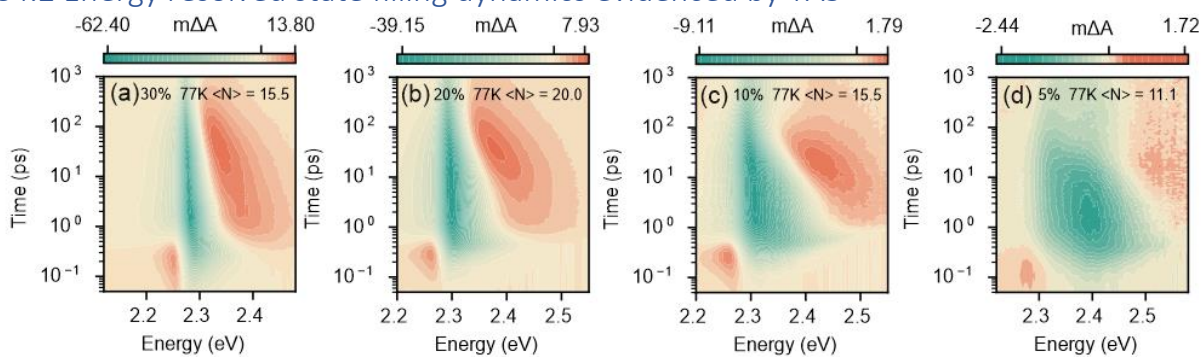

**Figure S4.2 Transient absorption colormaps for different samples with different $C_{prec}$.** False color maps of the change in absorption ($\Delta A$) at excitation intensities well above <N>=1 at 77K.